\documentclass[%
 reprint,
superscriptaddress,
nofootinbib,
 amsmath,amssymb,
 aps,
]{revtex4-2}

\usepackage{graphicx}
\usepackage{dcolumn}
\usepackage{bm}
\usepackage[colorlinks,linkcolor=red,anchorcolor=blue,citecolor=green]{hyperref}
\hypersetup{
    colorlinks=true,
    linkcolor=red,
    filecolor=gray,
    urlcolor=blue,
    citecolor=blue,
}

\usepackage{subfigure}
\usepackage{float}
\usepackage{graphicx}
\usepackage{booktabs}
\usepackage{tabularx}
\usepackage{comment}
\usepackage{microtype}
\usepackage{orcidlink}
\usepackage{pifont} %

\begin{document}
\preprint{APS/123-QED}

\title{
Imprints of core/cusp dark matter distributions on black hole signatures in galaxies
}

\author{Hassan Hassanabadi\,\orcidlink{0000-0001-7487-6898}
}
\email{hassanhassanabadi@mail.fresnostate.edu}
\affiliation{ Physics Department, California State University, Fresno, CA 93740, USA}

\author{Che-Yu Chen\,\orcidlink{0000-0001-8312-759X}
}
\email{b97202056@gmail.com}
\affiliation{RIKEN iTHEMS, Wako, Saitama 351-0198, Japan}

\author{Soroush Zare\,\orcidlink{0000-0003-0748-3386}
}
\email{soroushzrg@gmail.com}
\affiliation{ Helsinki Institute of Physics, University of Helsinki, P.O. Box 64, FI-00014 Helsinki, Finland}

\author{Volker Perlick\,\orcidlink{0000-0002-8400-8901}
}
\email{perlick@uni-bremen.de}
\affiliation{Faculty 1, University of Bremen, 28359 Bremen, Germany}


\begin{abstract}

In galactic environments, a host dark matter (DM) halo can imprint weak but coherent corrections on black hole (BH) strong-field observables.
We construct an exact family of static and spherically symmetric BH spacetimes sourced by a generic core/cusp DM halo, described by an anisotropic stress-energy tensor with nonvanishing radial pressure. 
The resulting geometry is determined self-consistently from the Einstein equations for a broad $\{\alpha,\beta,\gamma\}$ density profile, including the NFW, Moore, Hernquist, Jaffe, and core/cusp Dehnen models as special cases. 
We discuss the asymptotic structure, horizon location, curvature scale, and energy conditions of the corresponding geometries, emphasizing that the inner logarithmic slope $\gamma$ controls the amount of DM probed by the relativistic region.
We then obtain perturbative analytic estimates for the characteristic circular geodesics, demonstrating that the leading strong-field corrections are controlled by the dimensionless compactness $q_\gamma$, rather than by the total halo mass alone. 
To leading order in $q_\gamma$, we derive analytic expressions for the light ring radius, angular frequency, critical impact parameter, Lyapunov exponent, innermost
stable circular orbit (ISCO) radius, and ISCO frequency. For cuspy profiles, the light ring and ISCO are displaced outward, while the corresponding orbital frequencies are redshifted; for cored profiles, the light ring radius is unchanged at this order, although its frequency and capture impact parameter still carry finite environmental corrections. We also investigate the weak- and strong-deflection angles and their dependence on the halo’s inner structure. We further discuss how these environmental corrections may affect ringdown physics, in particular through the perturbative imprint of the halo on quasinormal-mode redshifts and late-time wave propagation. Our results provide a unified framework for assessing how galactic DM can leave controlled, slope-dependent signatures in the strong-field region of astrophysical BHs.

\end{abstract}

\maketitle


\section{Introduction}\label{sec1}

Black holes (BHs) in our Universe are rarely in isolation. Instead, they are usually accompanied by various astrophysical objects such as accretion disks, companion stars, or galactic halos \cite{KormendyARAA2013,BaraussePRD2014,BensonPR2010}. A faithful assessment of galactic environmental effects requires moving beyond order-of-magnitude estimates and constructing a self-consistent relativistic description of the system \cite{CardosoPRD2022,CardosoPRL2022,SpeeneyPRD2024,PezzellaPRD2025}. In particular, quantifying how the host galaxy influences gravitational dynamics, radiation generation and propagation, and the associated optical phenomena calls for a spacetime geometry describing a central BH embedded in an extended matter distribution \cite{CardosoPRD2022,CardosoPRL2022,SpeeneyPRD2024,PezzellaPRD2025,StuchlikApJ2022,KonoplyaPRD2025,KonoplyaApJ2022,XavierPRD2023,PatraJCAP2025,FonsecaPRD2026}. 
Within a static and approximately spherical framework, this surrounding matter can be modeled as an anisotropic fluid with a prescribed density profile, representing a galactic halo whose mass content is predominantly composed of dark matter (DM) \cite{CardosoPRD2022,CardosoPRL2022,SpeeneyPRD2024,PezzellaPRD2025}. 
Modeling compact objects in nonvacuum environments provides valuable insight into the role of astrophysical surroundings and their associated phenomenology in gravitational wave astronomy \cite{BertoneNat2018,CardosoLRR2019,MacedoApJ2013,EdaPRL2013, CardosoAA2020,KavanaghPRD2020}. 
A reliable description of galactic density profiles is informed by both observational data and large-scale numerical simulations.
As integral components of galaxies, DM halos play a central role in shaping the dynamical behavior of embedded objects and are usually described through a variety of density profiles \cite{HaroonPRD2025}, depending on the galaxy’s mass, size, and morphological properties \cite{Taylor2003,DehnenMNRAS1993,HernquistApJ1990,MooreMNRAS1999, NFWApJ1996,JaffeMNRAS1983,KingApJ1962}.

In this work, we consider a generic density distribution for a galactic DM halo, given by \cite{ZhaoMNRAS1996,ZhaoMNRAS1997}
\begin{equation}\label{DMDensityProfile}
	\rho(r) = \frac{(3-\gamma){\rm M}_{\rm DM}^{\rm tot}}{4\pi r_{\rm s}^{3}}\left(\frac{r}{r_{\rm s}}\right)^{-\gamma}\left[1+\left(\frac{r}{r_{\rm s}}\right)^{\alpha}\right]^{\frac{\gamma-\beta}{\alpha}}\,,
\end{equation}
where $r_{\rm s}$ denotes the scale radius, ${\rm M}_{\rm DM}^{\rm tot}$ is the total DM mass of the halo, and $\alpha$, $\beta$, and $\gamma$ are dimensionless parameters that characterize the specific density profile. These parameters encode the size, mass, and structural form of a galaxy and play a central role in determining the gravitational potential in the vicinity of supermassive BHs.
Adopting this density profile as the source term in the Einstein equations, and following the approach proposed in Ref. \cite{MaEPJC2024}, we obtain an exact and general solution describing the spacetime of a central BH immersed in a galactic halo, treated self-consistently through a solution of the Einstein equations sourced by an anisotropic stress-energy tensor 
$T^{\mu}{}_{\nu} = \mathrm{diag}(-\rho, P_{r}, P_{t}, P_{t}),$
where $\rho(r)$ is the halo density, and $P_{r}(r)$ and $P_{t}(r)$ denote the radial and tangential pressures, respectively.
For this general solution, the mass function is constructed by integrating the halo density profile $\rho(r)$, and the Einstein equations are solved under the conditions
$P_r(r) = -\rho(r)$, $P_t(r) = -\frac{r}{2}\rho'(r) - \rho(r)$.
Thus, the spacetime metric $g_{\mu\nu}$ is fully determined by the Einstein equations in the form of a static, spherically symmetric geometry, with $g_{tt}g_{rr} = -1$. The DM-induced correction to the radial metric function encodes the gravitational imprint of the surrounding DM halo on the spacetime geometry \cite{ShenPLB2025}.

With appropriate choices of the model parameters ${\alpha,\beta,\gamma}$, the generic density profile reproduces some widely used and well-tested DM halo models that are of interest in this study. In particular, well-known profiles such as the Navarro-Frenk-White (NFW) \cite{NFWApJ1996}, Moore \cite{MooreMNRAS1999}, Jaffe \cite{JaffeMNRAS1983}, and Hernquist \cite{HernquistApJ1990} models can be obtained as special cases and are known to provide good fits to numerical simulations of DM halos.
There is also another well-established and versatile class of models that can be obtained from the same generic form profiles, namely the core/cusp Dehnen density profiles by setting $(\alpha, \beta, \gamma) = (1, 4, \gamma)$, where $\gamma\in[0,3)$, controls the inner slope of the density distribution and uniquely specifies the particular variant of the profile, allowing one to interpolate continuously between cored and cuspy density distributions.

Dehnen-$(1,4,\gamma)$ profiles (a family of density distributions distinguished by their inner slopes), along with other closely related models \cite{DehnenMNRAS1993,HernquistApJ1990,MooreMNRAS1999,NFWApJ1996,JaffeMNRAS1983,KingApJ1962,ZhaoMNRAS1996}, typically exhibit enhanced densities toward galactic centers.
However, when a BH is present at the core, both Newtonian and relativistic analyses indicate that the DM density may get suppressed \cite{CardosoPRD2022,SadeghianPRD2013,GondoloPRL1999}. In this case, the DM distribution develops a cusp whose characteristic length scale is set by the BH mass $M_{\mathrm{BH}}$, thus the total mass contained in the cusp region is negligibly small.

With the BH-DM geometry in hand, we then investigate how the core/cusp structure of the halo modifies the characteristic null and timelike circular geodesics of the central BH.
In particular, we determine the light ring and ISCO, together with their orbital frequencies, the photon-capture impact parameter, and the instability rate of the null circular orbit. Our analysis shows that the leading strong-field corrections are governed primarily by the DM enclosed on BH scales and hence by the inner logarithmic slope of the halo, rather than by its total mass alone. Cuspy profiles generally displace the characteristic orbits outward, redshift their frequencies, enlarge the photon-capture scale, and weaken the radial instability of the light ring, whereas cored distributions leave a more strongly suppressed imprint. These correlated shifts provide a direct link between the central structure of galactic DM halos and potentially observable signatures of the BH geometry.

Here, gravitational lensing offers a complementary probe of both the relativistic geometry of a BH and the surrounding DM distribution, since photon trajectories encode the combined influence of the compact object and its galactic environment \cite{KonoplyaPLB795,PerlickPRD2022,CunhaGRG2018,PaulaPRD2026,BozzaGRG2001,BozzaPRD2002,VegettiSSR2024,GaoEPJC2023}. In the weak-deflection regime, light rays remain far from the unstable photon sphere and acquire small, profile-dependent corrections integrated along their paths, whereas near the critical impact parameter the deflection angle grows logarithmically and produces relativistic images governed by the photon-sphere geometry and the strong-deflection coefficients \cite{BozzaGRG2001,BozzaPRD2002,VegettiSSR2024,GaoEPJC2023,JuniorPRD2024,JhaJCAP2025,MollaPoDU2025}. Applying both limits to our self-consistent family of core/cusp BH-DM spacetimes, we find that the lensing signatures are controlled primarily by the inner logarithmic slope: weak-deflection corrections are strongly suppressed for cored profiles and become larger for steeper cusps, while, within the perturbative regime considered here, cuspy halos also increase the critical impact parameter and enhance the leading strong-deflection coefficient, which remains unchanged for the cored Dehnen profile at this order.

Beyond geodesic observables, the ringdown spectrum provides a further probe of the halo geometry. Previous studies of BHs surrounded by spherical matter distributions found that the environment generally lowers both the oscillation and damping rates of quasinormal modes, often through a common leading redshift \cite{CardosoPRD2022,PezzellaPRD2025,Chen:2023akf}. In our core/cusp spacetimes, however, the light-ring frequency and Lyapunov exponent acquire distinct corrections governed by the inner slope $\gamma$, with the scalar modes approaching the corresponding profile-dependent trajectory in the complex-frequency plane at large multipole number. Ringdown spectroscopy may therefore probe not only the strength of the galactic environment, but also the central structure of the DM halo.

The remainder of this paper is organized as follows. In Sec. \ref{sec2}, we construct the exact static and spherically symmetric BH geometries sourced by the generic $\{\alpha,\beta,\gamma\}$ DM distribution, examine their asymptotic and near-horizon properties, fluid interpretation, and energy conditions, and present explicit solutions for the halo profiles considered. Section \ref{sec3} is devoted to the null and timelike geodesic structure, including the profile-dependent corrections to the light ring, the ISCO, their associated frequencies, the photon-capture impact parameter, and the light-ring instability, together with the weak- and strong-deflection limits of gravitational lensing. In Sec. \ref{sec4}, we investigate the environmental shifts of the scalar quasinormal-mode spectrum and their connection with the light-ring dynamics in the eikonal regime. Finally, Sec. \ref{Conc} summarizes our main findings and discusses their phenomenological implications and possible extensions. 
Throughout this work, we adopt geometrized units in which $G=c=1$.

\section{Construction of black hole geometry in galactic center}\label{sec2}

\subsection{Setup and geometry}
We begin by constructing a general and exact solution describing the spacetime of a central BH immersed in a galactic halo, assuming a static and spherically symmetric metric of the form
\begin{eqnarray}\label{GeneralMetricForm}
	ds^{2}&\equiv& g_{\mu\nu}dx^{\mu}dx^{\nu}\\ \nonumber
	&=& -A(r)dt^{2}+B(r)^{-1}dr^{2}+C(r) (d\theta^{2}+\sin^{2}\theta d\varphi^{2})\,. 
\end{eqnarray}
In the most general setting, the metric coefficient functions $A(r)$, $B(r)$, and $C(r)$ are independent. In this work, however, we impose $A(r)\!=\! B(r)$ to simplify the analysis, and fix $C(r)=r^{2}$.  We further require the spacetime to be asymptotically flat, such that in the limit $r \to +\infty$ the metric functions satisfy $A(r)\!=\! B(r) \!=\! 1$.
We assume that Eq.~\eqref{GeneralMetricForm} satisfies the Einstein field equations for an anisotropic matter source, characterized by an energy density $\rho(r)$, a radial pressure $P_{r}(r)$, and a tangential pressure $P_{t}(r)$, with the corresponding stress-energy tensor given by
\begin{eqnarray}
8\pi T^{t}_{\,\,\,t} &=& -8\pi \rho = \frac{A(r)-1}{r^{2}}+\frac{A'(r)}{r}\,,\label{ttt}\\
8\pi T^{r}_{\,\,\,r} &=& 8\pi P_{r}(r)= -8\pi \rho\,, \\
8\pi T^{\theta}_{\,\,\,\theta} &=& 8\pi T^{\varphi}_{\,\,\,\varphi} = 8\pi P_{t}(r)= \frac{A'(r)}{r}+\frac{A''(r)}{2}\,,
\end{eqnarray}
where the prime stands for the derivative with respect to $r$. To specify the metric functions, we follow the approach proposed in Ref. \cite{MaEPJC2024}. In this framework, upon integrating the DM density distribution given in Eq.~\eqref{DMDensityProfile} and solving the Einstein field equations, the metric function can be obtained as
\begin{eqnarray}\label{metric function}
A(r) &=& 1-\frac{2M(r)}{r}\,,
\end{eqnarray}
where 
\begin{eqnarray}\label{Massfunction}
	M(r) &=& M_{\rm BH} + M_{\rm DM}(r)\,,\\
	\nonumber
	M_{\rm DM}(r) &=& {\rm M}_{\rm DM}^{\rm tot} \left(\frac{r}{r_{\rm s}}\right)^{3-\gamma}\\ \!&{}&_2F_1\!\left(\!\frac{3-\gamma}{\alpha}\!,\frac{\beta-\gamma}{\alpha};\frac{3-\gamma}{\alpha}+1;-\left(\frac{r}{r_{\rm s}}\right)^{\alpha}\right)\,,\nonumber\\
\end{eqnarray}
and $M_{\rm BH}$ arises as an integration constant which can be interpreted as BH mass. Note that according to Eq.~\eqref{ttt}, the mass function and the energy density are related by $M'(r)=4\pi\rho r^2$. Throughout this paper, we will assume $\alpha>0$.

The corresponding density profile for $\alpha>0$ exhibits an inner behavior $\rho(r)\propto r^{-\gamma}$ in the central region ($r\ll r_{\rm s}$) 
while at large distances ($r\gg r_{\rm s}$) it falls off as $\rho(r)\propto r^{-\beta}$ with parameters $\gamma$ and $\beta$ controlling the inner and outer slopes of the halo distribution, respectively. The smoothness of the transition between the inner and outer regimes is attributed to the parameter $\alpha$, which, in turn, controls how quickly the profile transitions from the inner to the outer region \cite{BaesMNRAS2021}.

In turn, models associated with varying $\gamma$ (restricted to the interval $[0, 3)$) are primarily distinct in their heart regions, extending to a few scale radii where the transition between $r^{-\gamma}$ and $r^{-\beta}$ transpires. However, these models may still exhibit considerable similarity in the envelope. The variation of the inner logarithmic slope $\gamma$ naturally divides the family into two regimes: $0\leq\gamma<1$, corresponding to cored or weakly cusped central profiles, and $1\leq\gamma<3$, describing more strongly cusped profiles, with $\gamma=1$ marking the transition between them \cite{HernquistApJ1990}.

Correspondingly, the enclosed DM mass behaves as
$M_{\rm DM}(r)\propto r^{3-\gamma}$ near the center, while at large radii it follows
\begin{eqnarray}
	M_{\rm DM}(r)\propto
	\begin{cases}
		\ln(r/r_s)\,, & \beta=3\,,\\[2mm]
		M_{\rm DM}(\infty)-r^{\frac{3-\beta}\alpha}\,, & \beta>3\,,
	\end{cases}
\end{eqnarray}
showing that $\beta=3$ separates infinite-mass halo configurations from finite-mass distributions.
In particular, only for $\beta>3$ does the quantity $M_{\rm DM}(\infty)$ remain finite, reflecting a sufficiently rapid decay of the matter distribution at large distances. Note that we have defined
\begin{eqnarray}
	M_{\rm DM} (\infty) = 4\pi \int^{\infty}_{0} 	\rho(r)r^{2} dr\,.
\end{eqnarray}
 For the generic mass profile, the quantity $M_{\rm DM} (\infty)$ and the total DM mass ${\rm M}_{\rm DM}^{\rm tot}$ are related by
\begin{eqnarray}
\!\!\!\!\!\!\!\!\!\!\!\!	M_{\rm DM} (\infty)\! =\! \frac{3-\gamma}{\alpha} B\left(\frac{3-\gamma}{\alpha},\frac{\beta-3}{\alpha}\right){\rm M}_{\rm DM}^{\rm tot}\!\!, \quad  \beta>3\,,
\end{eqnarray}
where $B(x,y)$ is the Euler beta function. In this case, the metric function $A(r)$ at large radii reads
\begin{eqnarray}
A(r) = 1-\frac{2}{r}\Big(M_{\rm BH}+{\rm M}_{\rm DM}^{\rm tot} C_{\alpha\beta\gamma}\Big)+\mathcal{O}\left(\frac1{r^2}\right)\,,
\end{eqnarray}
where
\begin{equation}
C_{\alpha\beta\gamma}\equiv \frac{3-\gamma}{\alpha} B\left(\frac{3-\gamma}{\alpha},\frac{\beta-3}{\alpha}\right)\,.
\end{equation}
Therefore the ADM mass read from the $1/r$ coefficient is $M_{\rm ADM}=M_{\rm BH}+{\rm M}_{\rm DM}^{\rm tot} C_{\alpha\beta\gamma}$. For special choices such as $\beta=\alpha+3$, which include the Dehnen family $\left(\alpha,\beta\right) = (1,4)$, the above relations reduce to $C_{\alpha\beta\gamma}=1$ and $M_{\rm DM} (\infty) = {\rm M}_{\rm DM}^{\rm tot}$, such that the ADM mass is given by $M_{\rm ADM}=M_{\rm BH}+{\rm M}_{\rm DM}^{\rm tot}$.

Obviously, the ADM mass is finite only for $\beta >3$, while for $\beta=3$, the mass function grows logarithmically and $M_{\rm ADM}$ is not well defined. However, the fact that it grows only logarithmically implies that the entire DM term in the metric function $A(r)$ still decays as $(\ln r)/r$. In both cases $\beta\ge 3$, the halo contribution to the metric vanishes at infinity and thus $A(r)\to 1$, so the spacetime is asymptotically flat.

\begin{figure}[H]
	\center{
		\includegraphics[width=8.5cm]{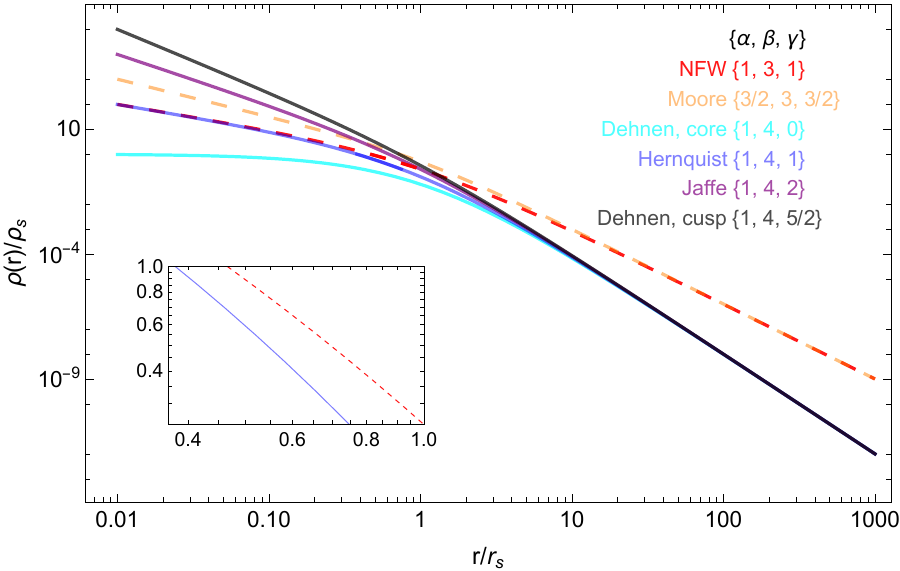}}
	\caption{The DM density $\rho(r)/\rho_{\rm s}$ is shown as a function of the radial coordinate $r/r_{\rm s}$, evaluated in the vicinity of the galactic center for representative choices within the $\{\alpha,\beta,\gamma\}$ family.
        }
	\label{densityProfiles}
\end{figure}

In Fig. \ref{densityProfiles}, we display the behavior of the density profiles $\rho(r)$ in the vicinity of the galactic center as a function of the radial coordinate $r$, adopting the scale density $\rho_{\rm s}=(3-\gamma) {\rm M}_{\rm DM}^{\rm tot}/4\pi r_{\rm s}^{3}$ for representative $\{\alpha,\beta,\gamma\}$-models. Shown are several widely used density profiles, including the NFW-$\{1,3,1\}$, Moore-$\{3/2,3,3/2\}$, cored Dehnen-$\{1,4,0\}$, Hernquist-$\{1,4,1\}$, Jaffe-$\{1,4,2\}$, and cuspy Dehnen-$\{1,4,5/2\}$ profiles.

Next, we discuss the spacetime near the BH horizon. One can show that, at the BH scale $r\sim 2M_{\rm BH}\ll r_{\rm s}$ the DM contribution to the lapse function behaves parametrically like
\begin{eqnarray}
\delta A(r)\sim -\frac{2M_{\rm DM}(r)}{r}\propto -r^{2-\gamma}\,,\\
M_{\rm DM}(2M_{\rm BH})\sim {\rm M}_{\rm DM}^{\rm tot}\Big(\frac{2M_{\rm BH}}{r_{\rm s}}\Big)^{3-\gamma}\,.\label{smallrmass}
\end{eqnarray}
Therefore, under a hierarchy $M_{\rm BH}\ll {\rm M}_{\rm DM}^{\rm tot}\ll r_{\rm s}$, 
the event horizon remains perturbatively close to the Schwarzschild value. Writing the horizon radius as $r_h = 2M_{\rm BH}+\delta r_{h}$, the leading correction induced by the surrounding DM halo is $\delta r_{h} \simeq 2M_{\rm DM}(2M_{\rm BH})$ provided that the enclosed DM mass near the horizon satisfies $M_{\rm DM}(2M_{\rm BH})\ll M_{\rm BH}$. 
Although the total halo mass may be much larger than the BH mass, the near-horizon correction is suppressed by the small scale ratio $2M_{\rm BH}/r_{\rm s}$. This suppression is stronger for shallower inner profiles, corresponding to smaller $\gamma$, and weakens as $\gamma \to 3$. The spacetime nevertheless retains a curvature singularity at $r = 0$, since the Schwarzschild-like central term $2M_{\rm BH}/r$ is not regularized.

The tangential pressure is regular at the horizon. Indeed, for $r\sim r_h\simeq2M_{\rm BH}\ll r_{\rm s}$, the density behaves as \(\rho\propto r^{-\gamma}\), and therefore \(P_t=-\rho-r\rho'/2\simeq(\gamma/2-1)\rho\), which is finite at any nonzero horizon radius.
The corresponding
near-horizon curvature scale induced by the halo is
\begin{eqnarray}
	R(r_h)\sim \rho(r_h)
	\sim
	\frac{{\rm M}_{\rm DM}^{\rm tot}}
	{r_{\rm s}^{3-\gamma}M_{\rm BH}^{\gamma}}\,,
\end{eqnarray}
so that, for a Hernquist-like cusp with $\gamma=1$, one obtains
\begin{eqnarray}
	R(r_h)\sim
	\frac{{\rm M}_{\rm DM}^{\rm tot}}{r_{\rm s}^2M_{\rm BH}}\,.
\end{eqnarray}
The corresponding dimensionless curvature is controlled by
\begin{eqnarray}
	R(r_h)M_{\rm BH}^2
	\sim
	\frac{{\rm M}_{\rm DM}^{\rm tot}}{M_{\rm BH}}
	\left(\frac{M_{\rm BH}}{r_{\rm s}}\right)^{3-\gamma}\,,
\end{eqnarray}
which can be made parametrically small under the hierarchy
$M_{\rm BH}\ll {\rm M}_{\rm DM}^{\rm tot}\ll r_{\rm s}$, provided the small scale ratio $M_{\rm BH}/r_{\rm s}$ dominates over the large mass ratio
$M_{\rm DM}^{\rm tot}/M_{\rm BH}$.

At large distances, for the particular anisotropic source adopted here, the exact
Ricci scalar is $R=8\pi(4\rho+r\rho')$.
Thus, for $\beta\neq4$ one obtains
\begin{eqnarray}
	R(r)\simeq
	2(3-\gamma)(4-\beta){\rm M}_{\rm DM}^{\rm tot}
	\frac{r_{\rm s}^{\beta-3}}{r^\beta}\,,
\end{eqnarray}
whereas the case $\beta=4$ is special: the leading $r^{-4}$ contribution
cancels, and the Ricci scalar decays faster,
\begin{eqnarray}
	R(r)\simeq
	2(3-\gamma)(4-\gamma){\rm M}_{\rm DM}^{\rm tot}
	\frac{r_{\rm s}^{1+\alpha}}{r^{4+\alpha}}\,.
\end{eqnarray}
In both cases, the behaviors of the Ricci scalar are consistent with the asymptotically flat nature of the spacetime.

\subsection{Fluid model}
As we have mentioned, for a metric of the form given in Eq.~\eqref{GeneralMetricForm} where $A(r)=B(r)$ and $C(r)=r^2$, the Einstein field equations give an energy-momentum tensor of the form $T^{\mu}{}_{\nu} = \mathrm{diag} \big( - \rho , P_r = - \rho , P_t , P_t \big)$, which stands for the energy-momentum tensor of a fluid with an anisotropic pressure. If the energy density $\rho$ is positive, the radial pressure $P_r$ is negative. Because of the degeneracy ($T^t{}_t = T^r{}_r$), the four-velocity of the fluid is not unique. One may consider any linear combination of $\partial _t$ and $\partial _r$ that is normalised as this four-velocity, i.e., any  vector field of the form
\begin{equation}
U^{\mu} (r) = \sqrt{ \dfrac{1}{A(r)}+\dfrac{u(r)^2}{A(r)^2}} \, 
\delta ^{\mu}_t - u(r) \, \delta ^{\mu} _r \,.
\end{equation}
Here $u(r)$ is arbitrary, except for the condition that 
\begin{equation}
\dfrac{1}{A(r)}+\dfrac{u(r)^2}{A(r)^2 }> 0
\label{eq:u}
\end{equation}
on the considered domain. For every choice of $u(r)$, the energy density
in the rest system of the fluid is $\rho$ and the radial pressure is $- \rho$.
\\[0.2cm]
Here we are interested in a spacetime which has one non-degenerate
horizon, i.e., in the case that there is a value $r_h$ such that
\begin{equation}
\begin{matrix}
- \infty < A(r) < 0 \quad \mathrm{for} \quad 0 < r < r_{\rm h}\,,
\\[0.2cm]
0 < A(r) < \infty \quad \mathrm{for} \quad r_{\rm h} < r < \infty \,.
\end{matrix}
\end{equation}
Then, by (\ref{eq:u}),  it is impossible to choose $u(r) = 0$ on or below
the horizon, so the four-velocity necessarily has a non-zero radial 
component. In the case of a black hole (as opposed to a white hole), 
we must have $u(r)>0$ on and below the horizon, i.e., the matter must 
be radially infalling. This makes it necessary to assume that there is 
an infinite reservoir of matter at infinity. \\

\subsection{Energy Conditions}
For the effective anisotropic source associated with the metric function $A(r)=B(r)=1-2M(r)/r$, the radial pressure satisfies $P_r=-\rho$. Consequently, the weak energy condition is satisfied whenever the halo density is nonnegative ($\rho \ge 0$) and monotonically decreasing ($\rho' \le 0$), since $\rho+P_{r}=0$, and $\rho+P_{t}=-(r\rho'/2)\ge0$. The strong energy condition (SEC) is more restrictive and requires $P_t\ge0$, or equivalently $-r\rho'\ge 2\rho$.
For a local cusp $\rho\propto r^{-\gamma}$, this condition reduces to $\gamma\ge2$. Thus, in the near-horizon region $r_h\ll r_s$, shallow cusps with $0\le\gamma<2$ violate the SEC, $\gamma=2$ marginally saturates it, and steeper cusps with $2<\gamma<3$ can satisfy it. For NFW- or Hernquist-type inner profiles, $\gamma=1$, and therefore $P_t(r_h)\simeq-\rho(r_h)/2$, implying SEC violation near the horizon. This does signal that the effective source develops tangential tension in the innermost region.
The dominant energy condition is instead governed by $\rho\ge |P_t|$, or $0\le K(r)\le4$, where $K(r)=-d\ln\rho/d\ln r$, and is therefore not generically violated near the horizon for the usual range $0\le \gamma<3$.\\

\subsection{Exact solutions of the models}
Before ending this section, we now provide a set of analytical solutions related to the lapse function for various density profiles $\rho(r)$ such as cusp NFW-$\{1,3,1\}$, cusp Moore-$\{3/2,3,3/2\}$ and the core/cusp Dehnen subfamily, setting $(\alpha, \beta, \gamma) = (1, 4, \gamma)$, with $\gamma$ defining the particular variant of the profile. Some allowed forms of the Dehnen profiles can be derived by setting core Dehnen-($\gamma = 0$), cusp Hernquist-($\gamma = 1$), cusp Jaffe-($\gamma = 2$), and cusp Dehnen-($\gamma = 5/2$).\\

\paragraph*{\bf Model I:} 
The spacetime for the BH in the cusp NFW-$\{1,3,1\}$ DM model takes the following form
\begin{equation}\label{coreNFW}
	\begin{split}
		A(r) = 1\!-\frac{2M_{\rm BH}}{r}\!+\frac{4{\rm M}_{\rm DM}^{\rm tot}}{r} \!\left(\frac{r}{r+r_{\rm s}}-\ln\!\left[\frac{r+r_{\rm s}}{r_{\rm s}}\right]\right)\,.
	\end{split}
\end{equation}

\paragraph*{\bf Model II:} 
The BH spacetime with the cusp Moore-$\{3/2,3,3/2\}$ DM model becomes
\begin{equation}\label{Moore}
	\begin{split}
		A(r) = 1-\frac{2M_{\rm BH}}{r}-\frac{2 {\rm M}_{\rm DM}^{\rm tot} 
		}{r} \ln\left[1+\left(\frac{r}{r_{s}}\right)^{3/2}\right]\,.
	\end{split}
\end{equation}

\paragraph*{\bf Model III:} 
The solution for the BH in the core Dehnen-type density distribution by setting $\{1,4,0\}$ takes the form
\begin{equation}\label{coreMod1}
	\begin{split}
		A(r) = 1-\frac{2M_{\rm BH}}{r}-\frac{2{\rm M}_{\rm DM}^{\rm tot}r^{2}}{\left(r+r_{\rm s}\right)^{3}}\,.
	\end{split}
\end{equation}

\paragraph*{\bf Model IV:} 
In this model, we adopt $\{1,4,1\}$ corresponding to the Hernquist profile. We then obtain $A(r)$ as follows

\begin{equation}\label{cuspyMod2}
	A(r) = 1-\frac{2M_{\rm BH}}{r}-\frac{2{\rm M}_{\rm DM}^{\rm tot}r}{\left(r+r_{\rm s}\right)^{2}}\,.
\end{equation}

\paragraph*{\bf Model V:}  
Now, by setting $\{1,4,2\}$ associated with the Jaffe-type density distribution, we find $A(r)$ as 

\begin{equation}\label{cuspyMod3}
	\begin{split}
		A(r)  = 1-\frac{2M_{\rm BH}}{r}-\frac{2{\rm M}_{\rm DM}^{\rm tot}}{r+r_{\rm s}}\,.
	\end{split}
\end{equation}

\paragraph*{\bf Model VI:}  
By considering the cusp Dehnen-type density distribution with $\{1,4,5/2\}$, the metric function $A(r)$ takes the form 
\begin{equation}\label{cuspyMod4}
	\begin{split}
		A(r)  = 1-\frac{2M_{\rm BH}}{r}-\frac{2{\rm M}_{\rm DM}^{\rm tot}}{\sqrt{r\left(r+r_{\rm s}\right)}}\,.
	\end{split}
\end{equation}

\section{Geodesic structure of the spacetime}\label{sec3}
\subsection{effective potential}
We now turn to the geodesic structure of the BH spacetime surrounded by the DM halo.
To do so, we consider the Lagrangian for geodesic motion,
$\mathcal{L}=\frac{1}{2}g_{\mu\nu}u^\mu u^\nu$,
which, for the metric in Eq.~\eqref{GeneralMetricForm} with $A(r)=B(r)$ and $C(r)=r^2$, takes the form
\begin{equation}
\!\!\!2\mathcal{L}
	=-A(r)\dot{t}^{2}+B(r)^{-1}\dot{r}^{2}
	+r^2\dot{\theta}^{2}
	+r^2\sin^2\theta\,\dot{\varphi}^{2}
	=-\epsilon\,.
\end{equation}
Here, $\epsilon=0$ and $\epsilon=1$ correspond, respectively, to null and timelike geodesics, while the overdot denotes differentiation with respect to the affine parameter $\tau$, which coincides with the proper time for timelike geodesics.
Owing to the spherical symmetry of the spacetime, the geodesic motion can be restricted, without loss of generality, to the equatorial plane, $\theta=\pi/2$. In this plane, the Lagrangian allows one to identify the canonical momenta conjugate to the remaining spacetime coordinates. In particular, the staticity and spherical symmetry of the geometry lead to two conserved quantities,
\begin{eqnarray}
	p_t=-A(r)\dot{t}=-E\,,\label{energy} 
	\quad
	p_\varphi=r^2\dot{\varphi}=L\,,\label{angular}
\end{eqnarray}
where $E$ and $L$ denote, respectively, the conserved energy and angular momentum of the orbits.
To proceed, it is useful to rewrite the reduced Lagrangian in terms of the conserved quantities introduced above. One obtains
\begin{equation}
	2\mathcal{L}
	=\frac{\dot{r}^{2}}{A(r)}+\frac{L^2}{r^2}
	-\frac{E^2}{A(r)}
	=-\epsilon\,,
	\label{lagrangian}
\end{equation}
which can equivalently be expressed as
\begin{equation}
	\dot{r}^{2}+V_{\rm eff}(r)=E^2\,,
	\label{Geod}
\end{equation}
where the effective potential governing the radial motion is defined as
\begin{equation}
	V_{\rm eff}(r)=A(r)\left(\epsilon+\frac{L^2}{r^2}\right)\,.
	\label{eq}
\end{equation}
Equation~\eqref{Geod} therefore provides the radial equation of motion for both null $(\epsilon=0)$ and timelike $(\epsilon=1)$ geodesics in the given spherically symmetric background.

\subsection{Characteristic circular geodesics}
As in the Newtonian central-force problem, circular geodesics are
identified with stationary points of the radial effective potential.
A circular orbit of radius $r_c$ is therefore characterized by
\begin{equation}
	\dot r=0\,, \qquad V_{\rm eff}'(r_c)=0\,.
\end{equation}
The first condition gives the energy relation
\begin{equation}
	E_c^2
	=
	A(r_c)\left(\epsilon+\frac{L_c^2}{r_c^2}\right)\,,
	\label{ECirCondGeneric}
\end{equation}
The extremality condition $V_{\rm eff}'(r_c)=0$ then yields the general equation for circular geodesics,
\begin{equation}
	\epsilon r_c^3 A'(r_c)
	-
	L_c^2\left(2A(r_c)-r_c A'(r_c)\right)=0\,.
	\label{CirCondGeneric}
\end{equation}

For timelike circular geodesics $\epsilon=1$, Eq.~\eqref{CirCondGeneric} gives the angular momentum required to sustain a circular orbit at $r=r_c$,
\begin{equation}
	L_c^2=
	\frac{r_c^3 A'(r_c)}{2A(r_c)-r_c A'(r_c)}\,.
	\label{LtimelikeGeneric}
\end{equation}
Substitution of this expression into Eq.~\eqref{ECirCondGeneric} yields the corresponding energy \cite{CardosoPRD2009},
\begin{equation}
	E_c^2=
	\frac{2A(r_c)^2}{2A(r_c)-r_c A'(r_c)}\,.
	\label{EtimelikeGeneric}
\end{equation}

For null geodesics $\epsilon=0$, the circular-orbit condition must be treated directly from $V_{\rm eff}'(r_c)=0$. Since a nontrivial null orbit has $L\neq0$, one obtains \cite{Atkinson1965} (see also, e.g., Refs.~\cite{PerlickPRD2022,CunhaGRG2018})
\begin{equation}
	2A(r_{\rm LR})-r_{\rm LR}A'(r_{\rm LR})=0\,,
	\label{LRgeneric}
\end{equation}
where $r_{\rm LR}$ denotes the light ring radius. 
It should be emphasized
that, for a generic static and spherically symmetric geometry written in areal-radius coordinates, the location of a circular null orbit is controlled by the redshift function $A(r)$. The function $B(r)$ enters the radial kinetic term and therefore affects the radial dynamics and stability properties, but it does not appear explicitly in the algebraic condition for the existence of circular orbits.

Essentially, in the fully generic case, light rings are obtained from Eq.~ 
\eqref{LRgeneric}, while timelike circular geodesics are described by Eqs.~
\eqref{LtimelikeGeneric} and \eqref{EtimelikeGeneric}.

For a null circular geodesic, the energy and angular momentum are not separately fixed by the circularity condition; instead, their ratio is determined. Defining the impact parameter $b=L/E$ and using the light ring condition \eqref{LRgeneric}, one finds \cite{Atkinson1965} (cf. again, e.g., Refs.~\cite{PerlickPRD2022,CunhaGRG2018})
\begin{equation}
	b_{\rm crit}^2=\frac{r_{\rm LR}^2}{A(r_{\rm LR})}\,.
	\label{bLRgeneric}
\end{equation}
This is the critical impact parameter corresponding to the light ring.

The coordinate angular velocity of a circular orbit is
\begin{equation}
	\Omega
	\equiv
	\frac{\dot\varphi}{\dot t}
	=
	\frac{L}{E}\frac{A(r_c)}{r_c^{2}} \,.
\end{equation}
Using Eqs.~\eqref{LtimelikeGeneric} and \eqref{EtimelikeGeneric}, one obtains the
generic expression $	\Omega_c^{2}=A'(r_c)/2r_c$.
For null circular orbits, Eq.~\eqref{LRgeneric} gives
\begin{equation}
	\Omega_{\rm LR}
	=
	\frac{\sqrt{A(r_{\rm LR})}}{r_{\rm LR}}\,.
	\label{OmegaLRGeneric}
\end{equation}

Besides defining the photon-capture threshold, the critical impact parameter determines the angular radius of the BH shadow. For a static observer located at the areal radius $r_{\rm O}>r_{\rm LR}$, Eqs. (23) and (24) of Ref. \cite{PerlickPRD2022} yield
\begin{equation}
\alpha_{\rm sh} \!
=\! \arcsin\!\left[ \frac{b_{\rm crit}\sqrt{A(r_{\rm O})}}{r_{\rm O}} \right],
\end{equation}
where we have used 
\begin{equation}
    h(r)=r/\sqrt{A(r)}
\label{eq:h}
\end{equation} 
for the metric in Eq. \eqref{GeneralMetricForm} with $A(r)=B(r)$ and $C(r)=r^2$. For an observer sufficiently far from the BH, $A(r_{\rm O})\rightarrow1$ and $\alpha_{\rm sh}\ll1$, so that
\begin{equation}\label{AngShadowRad}
\alpha_{\rm sh}\simeq\frac{b_{\rm crit}}{r_{\rm O}}.
\end{equation}
Furthermore, comparison of Eqs. \eqref{bLRgeneric} and \eqref{OmegaLRGeneric} gives the exact relation
\begin{equation}\label{OmegaLRGeneric2}
\Omega_{\rm LR} = \frac{1}{b_{\rm crit}},
\end{equation}
in units with $c=1$. 
Thus, in the distant-observer limit and at a fixed observer radius, an increase in the critical impact parameter produces a larger angular shadow radius \eqref{AngShadowRad} while the exact relation \eqref{OmegaLRGeneric2} implies a corresponding decrease in the light ring angular frequency.

The stability of circular geodesics is determined by the second
derivative of the effective potential. For fixed \(E\) and \(L\), one
has
\begin{equation}
\!\!\!\!	V_{\rm eff}''(r)
	\!=\!
	A''(r)\left(\epsilon+\frac{L^{2}}{r^{2}}\right)
	-
	\frac{4A'(r)L^{2}}{r^{3}}
	+
	\frac{6A(r)L^{2}}{r^{4}}\,.
\end{equation}
A circular orbit in the static exterior region, where $A(r_c)>0$, is stable if $V_{\mathrm{eff}}''(r_c)>0$ and unstable if $V_{\mathrm{eff}}''(r_c)<0$. In the generic nondegenerate case, a marginally stable circular orbit satisfies $V_{\mathrm{eff}}''(r_c)=0$,
and
$ V_{\mathrm{eff}}'''(r_c)\neq0$.
If the third derivative also vanishes, the linear stability criterion is inconclusive, and the local behavior of the orbit must be determined from the first nonvanishing higher derivative of $V_{\mathrm{eff}}(r)$.
For timelike circular geodesics, substituting Eq.~\eqref{LtimelikeGeneric}
into the above expression yields
\begin{equation}
\!\!\!\!	V_{\rm eff}''(r_c)
	\!=\!
	\frac{
		2A(r) A''(r)
		-
		4(A'(r))^{2}
		+
		6A(r) A'(r)/r_c
	}{
		2A(r)-r_c A'(r)
	} \,.
	\label{VsecondTimelikeGeneric}
\end{equation}
Hence the marginally stable circular orbit is determined by \cite{Huang2026}
\begin{equation}
	A(r) A''(r)
	-
	2(A'(r))^{2}
	+
	\frac{3A(r) A'(r)}{r_c}
	=
	0\,.
	\label{MSCOGeneric}
\end{equation}
When several solutions exist, the smallest solution outside the
horizon and outside the light ring corresponds to the innermost
stable circular orbit (ISCO).

For null circular geodesics, using Eq.~\eqref{LRgeneric}
one finds
\begin{equation}
	V_{\rm eff}''(r_{\rm LR})
	=
	\frac{L^{2}}{r_{\rm LR}^{2}}
	\left[
	A''(r_{\rm LR})
	-
	\frac{2A(r_{\rm LR})}{r_{\rm LR}^{2}}
	\right]\,.
	\label{VsecondLRGeneric}
\end{equation}
Thus a light ring is unstable whenever the quantity in brackets is
negative, and stable whenever it is positive.

Another useful diagnostic of the geodesic structure is the Lyapunov exponent \cite{CardosoPRD2009,DeichPRD2024}, which quantifies the rate at which nearby trajectories separate from a given circular orbit and therefore characterizes the orbit’s dynamical instability.
For a static and spherically symmetric
metric with $A(r)=B(r)$, it is given by
\begin{equation}
	\lambda^{2}
	=
	-\frac{A(r_{c})^{2}}{2E_c^{2}}
	V_{\rm eff}''(r_c)\,.
	\label{LyapunovGeneric}
\end{equation}
Using Eq.~\eqref{VsecondLRGeneric}, the light ring Lyapunov exponent becomes
\begin{equation}
	\lambda_{\rm LR}
	=
	\Omega_{\rm LR}
	\left[
	A(r_{\rm LR})
	-
	\frac{	A''(r_{\rm LR})r_{\rm LR}^{2}}{2}
	\right]^{1/2}\,.
	\label{LyapunovLRGeneric3}
\end{equation}
A real positive value of \(\lambda_{\rm LR}\) corresponds to an
unstable light ring, while an imaginary value signals radial stability.
The light ring acts as a geometric imprint of the BH spacetime on the observed image. The frequency $\Omega_{\rm LR}$ is tied to the characteristic scale of the shadow boundary, whereas the Lyapunov exponent $\lambda_{\rm LR}$ measures the rate at which photons peel away from the unstable null orbit, setting the associated instability timescale $\lambda_{\rm LR}^{-1}$ and shaping the photon ring hierarchy. These two observables are therefore not independent probes: they reflect different facets of the same null geodesic structure, controlled by the underlying mass function.\\

In the Schwarzschild limit, where the halo contribution is switched off, the unstable light ring and the innermost stable circular orbit are located at
\(r_{\rm LR}=3M_{\rm BH}\) and \(r_{\rm ISCO}=6M_{\rm BH}\), respectively \cite{Chandrasekhar}.
When the central BH is embedded in the DM distribution, these circular geodesics are no longer those of a pure vacuum geometry. The halo modifies both their radial position and the corresponding coordinate frequencies. We shall focus on the galaxy regime in which the BH scale is much smaller than the halo scale, $M_{\rm BH}\ll r_{\rm s}$, and the matter contribution remains perturbative throughout the strong-field region. 

For the inner region of the density profile, \(r\ll r_{\rm s}\), the enclosed DM mass admits the leading small-radius scaling given by $M_{\rm DM}(r)\propto r^{3-\gamma}$. Consequently, the effective compactness of the halo in the strong-field region, $r\sim M_{\rm BH}$, is controlled by the dimensionless parameter
\begin{equation}
	q_\gamma
	=
	\frac{M_{\rm DM}^{\rm tot}M_{\rm BH}^{2-\gamma}}
	{r_{\rm s}^{3-\gamma}}
	=
	\frac{M_{\rm DM}^{\rm tot}}{r_{\rm s}}
	\left(\frac{M_{\rm BH}}{r_{\rm s}}\right)^{2-\gamma}\,.
	\label{qgamma_def}
\end{equation}
The perturbative regime considered below is therefore \(q_\gamma\ll1\). This condition states that, although the total halo mass may be much larger than the BH mass on galactic scales, the amount of matter enclosed inside the relativistic region is small compared with the central BH contribution. In this limit, the leading corrections are governed only by the inner slope \(\gamma\); the parameters \(\alpha\) and \(\beta\) enter at subleading order through corrections in powers of \(M_{\rm BH}/r_{\rm s}\). Keeping the first nonvanishing order in \(q_\gamma\), the light ring radius is shifted to
\begin{eqnarray}
	r_{\rm LR}
	\simeq
	3M_{\rm BH}
	\left[
	1+\gamma 3^{2-\gamma}q_\gamma
	\right]\,.
	\label{rLR_qgamma}
\end{eqnarray}
The corresponding angular frequency is
\begin{eqnarray}
	M_{\rm BH}\Omega_{\rm LR}
	\simeq
	\frac{1}{3\sqrt{3}}
	\left[
	1-3^{3-\gamma}q_\gamma
	\right]\,.
	\label{OmegaLR_qgamma}
\end{eqnarray}
The associated critical impact parameter for high-frequency photon capture is
\begin{eqnarray}
	b_{\rm crit}
	\simeq
	3\sqrt{3}M_{\rm BH}
	\left[
	1+3^{3-\gamma}q_\gamma
	\right]\,.
	\label{bcrit_qgamma}
\end{eqnarray}
Thus, at leading order, the halo increases the apparent critical impact parameter while redshifting the light ring frequency. This is the expected behavior for a weak but extended matter distribution surrounding the central object. 
Nevertheless, these environmental corrections are not expected to prevent light-ring-based tests of the nature of the central object from being carried out with good precision using the Event Horizon Telescope, GRAVITY, and comparable high-angular-resolution facilities \cite{EHT2019L1,GravityAA2020,GravityAA2018}.
The instability of the null circular orbit is characterized by the Lyapunov exponent \(\lambda\). To the same perturbative order, one finds
\begin{eqnarray}
	M_{\rm BH}\lambda
	\simeq
	\frac{1}{3\sqrt{3}}
	\left[
	1+
	\frac{1}{2}
	3^{2-\gamma}
	\left(
	\gamma^2-3\gamma-6
	\right)q_\gamma
	\right]\,.
	\label{lambdaLR_qgamma}
\end{eqnarray}
For \(0\leq\gamma<3\), the coefficient \(\gamma^2-3\gamma-6\) is negative, so the Lyapunov exponent decreases. Equivalently, the instability time scale \(\lambda^{-1}\) becomes longer in the presence of diffuse halos.

The timelike circular geodesics are modified in the same perturbative regime. The ISCO radius is displaced from its Schwarzschild value according to
\begin{eqnarray}
	r_{\rm ISCO}
	\simeq
	6M_{\rm BH}
	\left[
	1+
	2\,6^{2-\gamma}
	\left(
	2\gamma^2-8\gamma+9
	\right)q_\gamma
	\right]\,.
	\label{rISCO_qgamma}
\end{eqnarray}
The associated ISCO frequency becomes
\begin{eqnarray}
\!\!\!\! M_{\rm BH}\Omega_{\rm ISCO} \!
	\simeq \!
	\frac{1}{6\sqrt{6}}\!
	\left[
	1
	-
	\frac{1}{2}
	6^{3-\gamma}
	\left(
	2\gamma^2-9\gamma+11
	\right)q_\gamma
	\right]\!\!\,.
	\label{OmegaISCO_qgamma}
\end{eqnarray}

These analytic estimates show that, for cuspy profiles with $\gamma>0$, the light ring is displaced outward, while its angular frequency decreases and the critical impact parameter increases. For a cored profile with $\gamma=0$, the light ring radius remains unchanged at this order, although $\Omega_{\rm LR}$ and $b_{\rm crit}$ still receive finite corrections.

On the other hand, since both polynomials $2\gamma^2-8\gamma+9$ and $2\gamma^2-9\gamma+11$ are positive for $0\leq\gamma<3$, the leading halo correction pushes the ISCO outward and lowers the ISCO frequency. The Schwarzschild results are recovered smoothly in the limit $q_\gamma\to0$.

\subsection{Weak-deflection angle}
In the following subsections, we investigate the deflection angles of light rays propagating around our BH-DM system. 
For an asymptotically flat geometry, $A(r)\to 1$ as $r\to\infty$, the deflection angle of a photon arriving from infinity, reaching a radial turning point $r_0$, and escaping back to infinity \cite{Weinberg} can be written in the standard form
\begin{equation}
	\delta+\pi
	=
	2\int_{r_0}^{\infty}
	\left[
	\frac{h(r)^{2}}{h(r_0)^{2}}-1
	\right]^{-1/2}
	\frac{dr}{r\sqrt{A(r)}}\,.
	\label{BendingAngleGeneric}
\end{equation}
with the function $h(r)$ from (\ref{eq:h}).
Here $r_0$ denotes the radius of closest approach. Hence, the impact parameter of the ray is $b\,=\, h(r_{0})$. Thus, for the usual branch of photons outside the outermost unstable light ring, decreasing $r_0$ from infinity toward $r_{\rm LR}$ increases the bending angle from $\delta=0$ to a divergent value, while the impact parameter decreases from infinity to the critical impact parameter $b_{\rm crit} = h(r_{\rm LR})$.

To quantify the profile dependence of Eq.~\eqref{BendingAngleGeneric}, we first consider the weak-deflection regime \(M_{\rm BH}/r_0\ll1\) and introduce
\begin{eqnarray}
	\nonumber
&&\mu\equiv\frac{M_{\rm DM}^{\rm tot}}{r_{\rm s}}\,,
\qquad
x\equiv\frac{r}{r_{\rm s}}\,,
\qquad
x_0\equiv\frac{r_0}{r_{\rm s}}\,,\\
&&\mu f_i(x)\equiv A_i(r)-1+\frac{2M_{\rm BH}}{r}\,,
\end{eqnarray}
where the subscript $i$ labels the halo models. To leading order, the bending angle becomes
\begin{eqnarray}
\delta_i(r_0)
&\simeq&
\frac{4M_{\rm BH}}{r_0}+\mu\mathcal{F}_i(x_0)\,,\\
\mathcal{F}_i(x_0)&=&\int_0^{\pi/2}
\left[
x\frac{d f_i}{dx}-f_i
\right]_{x=x_0/\sin\vartheta}
d\vartheta \,,
\end{eqnarray}
where terms of order
\(\mathcal{O}[(M_{\rm BH}/r_0)^2,\mu M_{\rm BH}/r_0,\mu^2]\)
have been neglected.
	
In the region \(M_{\rm BH}\ll r_0\ll r_{\rm s}\), the leading halo
contributions are
\begin{eqnarray}
\delta_{\gamma=0}^{\rm DM}
&\simeq&
\mu x_0^3
\left[6\ln\left(\frac{2}{x_0}\right)-11\right]\,,
\nonumber\\
\delta_{\rm NFW}^{\rm DM}
&\simeq&
\delta_{\rm Hernquist}^{\rm DM}
\simeq2\mu x_0\,,
\nonumber\\
\delta_{\rm Moore}^{\rm DM}
&\simeq&
2.622\,\mu x_0^{1/2}\,,\\
\delta_{\rm Jaffe}^{\rm DM}
&\simeq&
\mu(\pi-2x_0)\,,
\nonumber\\
\delta_{\gamma=5/2}^{\rm DM}
&\simeq&
3.594\,\mu x_0^{-1/2}\nonumber\,.
\end{eqnarray}
Thus, at fixed \(\mu\) and \(x_0\ll1\), the halo correction follows
\begin{eqnarray}
\delta_{\gamma=5/2}^{\rm DM}
\!>\!
\delta_{\rm Jaffe}^{\rm DM}
\!>\!
\delta_{\rm Moore}^{\rm DM}
\!>\!
\delta_{\rm NFW}^{\rm DM}
\! \simeq \!
\delta_{\rm Hernquist}^{\rm DM}
\!>\!
\delta_{\gamma=0}^{\rm DM}\,.
\end{eqnarray}

One can see that the DM contribution to the bending angle is controlled primarily by the halo's inner structure. For a cored profile, the correction is strongly suppressed, whereas cuspy profiles produce increasingly larger deviations as the inner logarithmic slope grows. Thus, at fixed halo scale and total compactness, the deflection angle is enhanced monotonically with the cusp strength, showing that lensing in the relativistic region is sensitive mainly to the amount of DM enclosed near the BH rather than to the total halo mass alone.

\subsection{Strong-deflection angle}

The light ring quantities discussed above also determine the near-critical bending of photons. In contrast to the weak-deflection regime, where the closest approach satisfies $M_{\rm BH}/r_0\ll 1$, the strong-deflection regime is reached when the photon impact parameter approaches the critical value from above \cite{BozzaPRD2002}. It is useful to introduce the dimensionless distance from criticality
\begin{equation}
\epsilon_i\equiv \frac{b}{b_{{\rm crit},i}}-1\,,
\qquad 0<\epsilon_i\ll 1\,,
\label{eq:epsilon_i}
\end{equation}
where again the subscript $i$ labels the halo models.

The logarithmic form follows directly from the structure of the deflection integral near the unstable light ring. Defining
\begin{equation}
H(r)\equiv h^2(r)=\frac{r^2}{A(r)}\,,
\end{equation}
the light ring is a stationary point of $H(r)$, namely $H'(r_{\rm LR})=0$. We write the closest approach as
\begin{equation}
r_0=r_{\rm LR}+\Delta\,,
\qquad 0<\Delta\ll r_{\rm LR}\,,
\end{equation}
and introduce the local variable measured from the turning point,
\begin{equation}
z=r-r_0\,.
\end{equation}
Thus the lower limit of the bending integral is $z=0$, while
$r-r_{\rm LR}=z+\Delta$. Expanding $H(r)$ about the light ring gives
\begin{eqnarray}
\nonumber
H(r)-H(r_0)
&\simeq&
\frac{1}{2}H''_{\rm LR}
\left[(z+\Delta)^2-\Delta^2\right]\\
&\simeq&
\frac{1}{2}H''_{\rm LR}z(z+2\Delta)\,.
\label{eq:Hexpansion_turning}
\end{eqnarray}
The regular prefactors in the bending integral can be evaluated at
$r=r_{\rm LR}$ when extracting only the divergent part. Therefore the
near-critical contribution is controlled by \cite{BozzaPRD2002}
\begin{equation}
\int_0^{\ell_c}\frac{dz}{\sqrt{z(z+2\Delta)}}
=2\sinh^{-1}\sqrt{\frac{\ell_c}{2\Delta}}
\simeq
\ln\left(\frac{2\ell_c}{\Delta}\right)\,,
\label{eq:log_integral_z}
\end{equation}
where $\ell_c$ is a small fixed cutoff around the light ring. The remaining
part of the radial integral is finite and is absorbed into the regular
constant. Since $b^2=H(r_0)$ and $b_{\rm crit}^2=H(r_{\rm LR})$, one has
\begin{equation}
\frac{b}{b_{\rm crit}}-1
\simeq
\frac{H''_{\rm LR}}{4H_{\rm LR}}\Delta^2 \,.
\label{eq:b_delta_relation}
\end{equation}
Equations~\eqref{eq:log_integral_z} and \eqref{eq:b_delta_relation}
then imply the universal strong-deflection expansion
\begin{equation}
\delta_i(b)
=
-\bar a_i\ln\epsilon_i+\bar b_i
+{\cal O}(\epsilon_i\ln\epsilon_i)\,.
\label{eq:strong_deflection_general}
\end{equation}
Here $\bar a_i$ is fixed locally by the unstable light ring, whereas
$\bar b_i$ is a finite regular term depending on the full radial profile
of the metric function. For the class of geometries considered here,
$A(r)=B(r)$, the logarithmic coefficient may be written as
\begin{equation}
\bar a_i=
\sqrt{\frac{2}{2A_i(r_{{\rm LR},i})
-r_{{\rm LR},i}^{2}A_i''(r_{{\rm LR},i})}}
=
\frac{\Omega_{{\rm LR},i}}{\lambda_{{\rm LR},i}} \,.
\label{eq:abar_general}
\end{equation}
Thus the same null circular orbit that fixes the shadow scale and the
instability timescale also controls the leading strong-deflection
divergence.

Using the perturbative light-ring frequency and Lyapunov exponent obtained above (Eqs.~\eqref{OmegaLR_qgamma} and \eqref{lambdaLR_qgamma}), the leading halo correction to the logarithmic coefficient becomes
\begin{equation}
\bar a_\gamma
\simeq
1+\frac{1}{2}\,\gamma(3-\gamma)3^{2-\gamma}q_\gamma \,.
\label{eq:abar_gamma}
\end{equation}
The Schwarzschild value $\bar a=1$ is recovered when the halo contribution
is removed. For $0<\gamma<3$, the correction is positive; hence the halo
increases the coefficient of the logarithmic divergence. At the same
relative distance from the critical impact parameter, photons therefore
accumulate a larger bending angle in the presence of a cuspier halo.

\begin{table}[t]
\centering
\caption{The leading DM corrections to the logarithmic strong-deflection
coefficient for the six halo models, i.e., Eq.~\eqref{eq:abar_gamma}}
\begin{ruledtabular}
\begin{tabular}{lcc}
Model & $\gamma_{\rm inner}$ & $\bar a_i-1$ \\
\hline
I: NFW & $1$ & $3q_1$ \\
II: Moore & $3/2$ & $\dfrac{9\sqrt{3}}{8}q_{3/2}$ \\
III: core Dehnen & $0$ & $0$ \\
IV: Hernquist & $1$ & $3q_1$ \\
V: Jaffe & $2$ & $q_2$ \\
VI: cusp Dehnen & $5/2$ & $\dfrac{5}{8\sqrt{3}}q_{5/2}$ \\
\end{tabular}
\end{ruledtabular}
\label{tab:abar_dm_models}
\end{table}

The comparison among the six halo models is controlled, at the leading order, by the inner logarithmic slope according to Eq.~\eqref{eq:abar_gamma}, and is summarized in Table~\ref{tab:abar_dm_models}. This table should be interpreted together with the definition of
$q_\gamma$. At fixed halo compactness $M_{\rm DM}^{\rm tot}/r_{\rm s}$ and for
$M_{\rm BH}\ll r_{\rm s}$, $q_\gamma$ increases as the inner slope becomes larger.
Consequently, the strongest near-critical bending is produced by the
steepest cusps, while the cored profile gives the weakest effect.

The same conclusion follows from the critical impact parameter, since the
halo increases $b_{\rm crit}$ and shifts the photon-capture threshold
outward. Thus, for fixed $M_{\rm DM}^{\rm tot}/r_{\rm s}$ and
$M_{\rm BH}/r_{\rm s}\ll1$, the leading strong-field hierarchy is
\begin{equation}
\!\!\delta_{\gamma=5/2}
\!>\!
\delta_{\rm Jaffe}
\!>\!
\delta_{\rm Moore}
\!>\!
\delta_{\rm NFW}
\!\simeq\!
\delta_{\rm Hernquist}
\!>\!
\delta_{\gamma=0}\,,
\label{eq:strong_hierarchy}
\end{equation}
provided the comparison is made at the same relative distance
$\epsilon_i$ from each model's critical impact parameter and only the
leading logarithmic part is compared. The NFW and Hernquist models have
the same leading strong-field behavior because they share the same inner
slope, $\gamma=1$. They can still differ at subleading order and through
the finite constant $\bar b_i$, since $\bar b_i$ depends on the complete
radial form of $A_i(r)$. Similarly, the Moore and NFW cases should be
treated with care because their large-radius masses are logarithmically
divergent unless a cutoff is introduced; in this sense
$M_{\rm DM}^{\rm tot}$ acts as a profile normalization for these two
models.

The strong-deflection comparison is therefore complementary to the
weak-deflection one. The weak-deflection angle probes photons with
$r_0\gg M_{\rm BH}$ and is sensitive to the integrated halo correction
along a large-radius trajectory. The strong-deflection angle instead
probes photons that linger near the unstable light ring, and is therefore
controlled mainly by the amount of dark matter enclosed in the
relativistic region. This is why the inner cusp parameter $\gamma$ provides
the dominant ordering of the strong-field lensing imprint.

\section{Redshifts of quasinormal mode spectra}\label{sec4}

\begin{figure*}
  \centering
  \subfigure[]
{ \includegraphics[scale=0.45]{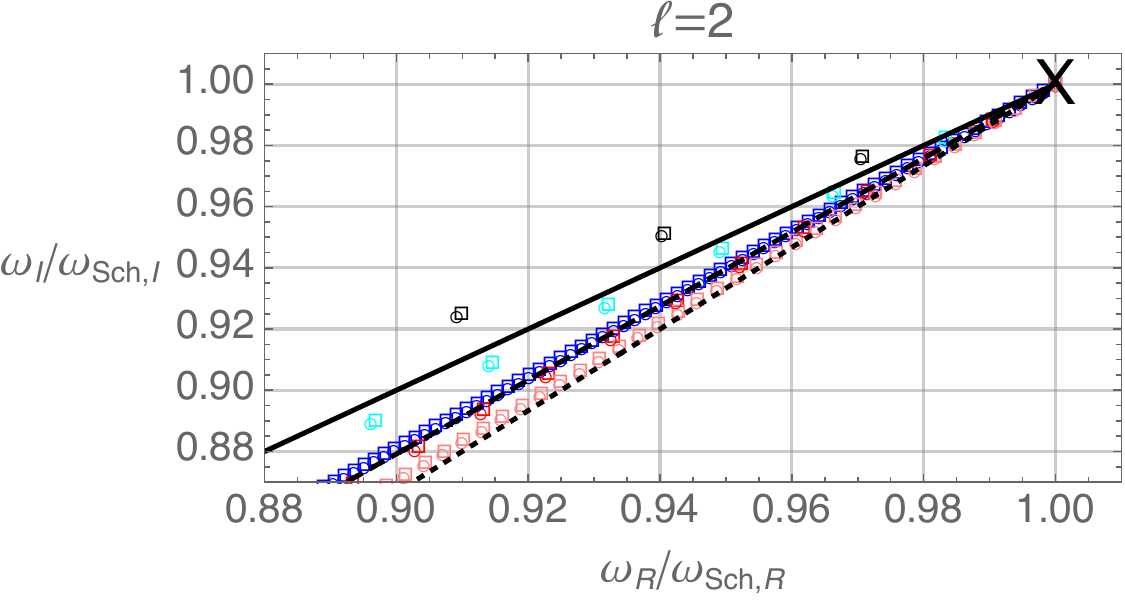}}
\subfigure[]
  {\includegraphics[scale=0.45]{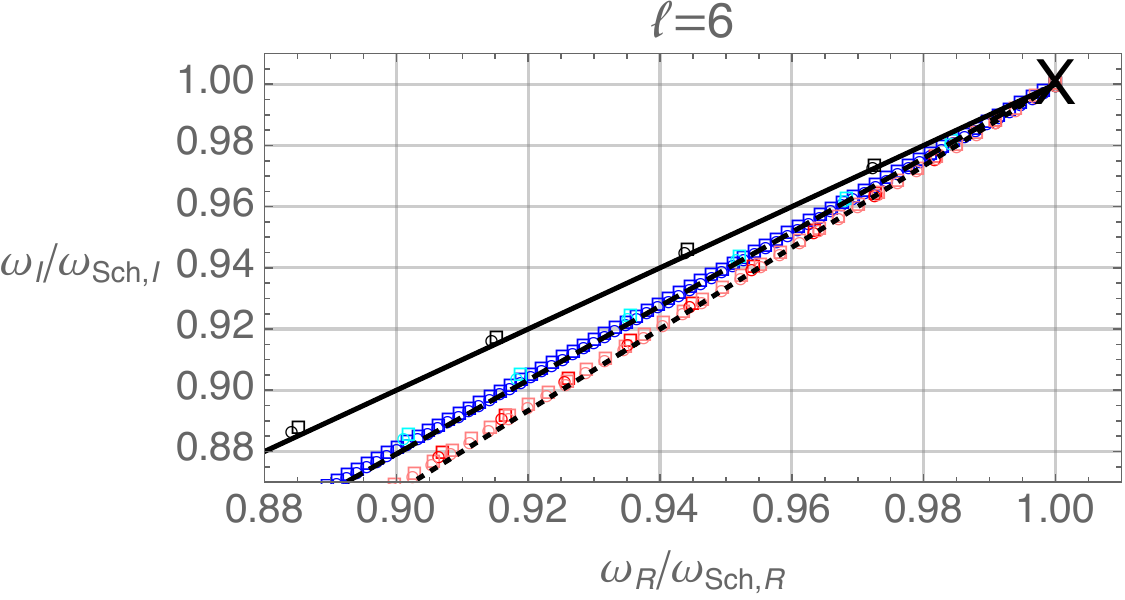}}
\\
\caption{The fundamental QNM frequency of massless scalar fields in the presence of a DM halo, with (a) $\ell=2$ and (b) $\ell=6$. We fix $r_{\rm s}=1000M_{\rm BH}$. The horizontal and vertical axes show the real and the imaginary parts of the frequencies relative to the Schwarzschild values. The cross on the right end represents the Schwarzschild case, i.e., $(1,1)$. The circles and squares represent $\alpha=1$ and $\alpha=2$, respectively. Here, we consider only $\beta=\alpha+3$. The black, cyan, red, pink, and blue points correspond to $\gamma=0$, $1/2$, $1$, $2$, and $5/2$, respectively. Each point extending from the cross labels an incremental increase of $q_\gamma$ by $1/1000$. The straight lines have slope $-(\gamma^2-3\gamma-6)/6$, with $\gamma=0$ (solid), $\gamma=(1/2,5/2)$ (dashed), and $\gamma=(1,2)$ (dotted). In the presence of DM halos, QNM frequencies are redshifted in a manner that, in the large-$\ell$ limit, only depends on $\gamma$.}
\label{fig:dmqnm} 
\end{figure*}

BH quasinormal modes (QNMs) characterize the natural frequencies of a perturbed BH and play a crucial role in the ringdown phase during binary BH mergers. In the presence of DM halos, the gravity induced by the halos shifts the QNM spectrum in a way that may depend on the DM profiles or on the setup of the BH-DM system. In Ref.~\cite{PezzellaPRD2025}, by assuming a zero radial pressure, it was shown that the QNM spectra are redshifted in the presence of spherically symmetric halos in a way that the real part and the imaginary part of the frequencies, relative to the Schwarzschild values, are shifted by a universal relation:{\footnote{A similar redshift relation has also been found in a system of BHs superposed with gravitating thin disks \cite{Chen:2023akf}.}}
\begin{equation}
\frac{\omega_R}{\omega_{\textrm{Sch,}R}}\approx\frac{\omega_I}{\omega_{\textrm{Sch,}I}}\approx1-\frac{M_{\rm DM}^{\rm tot}}{r_{\rm s}}+\mathcal{O}\left(\frac{M_{\rm DM}^{\rm tot}}{r_{\rm s}}\right)^2\,.
\end{equation}
This result can be explained using the well-known correspondence between unstable circular photon orbits and the high-frequency QNMs \cite{CardosoPRD2009}, i.e., 
\begin{equation}
\omega_R\approx \ell \Omega_\textrm{LR}\,,\quad \omega_I\approx-\frac{1}{2}|\lambda_\textrm{LR}|\,.
\end{equation}
where $\ell$ is the multipole number of QNMs. Note that here we have only considered fundamental modes. In the setup of Ref.~\cite{PezzellaPRD2025}, one can explicitly show that
\begin{equation}
M_{\rm BH}\Omega_{\rm LR}\approx M_{\rm BH}\lambda\approx\frac{1}{3\sqrt{3}}\left(1-\frac{M_{\rm DM}^{\rm tot}}{r_{\rm s}}\right)\,,
\end{equation}
at the leading order of DM corrections.

However, in our setup, the light ring angular frequency and the Lyapunov exponent are shifted in different ways that depend on $\gamma$, i.e., Eqs.~\eqref{OmegaLR_qgamma} and \eqref{lambdaLR_qgamma}. Indeed, in the presence of DM halos, the QNM frequencies are redshifted, but in a different way from that of Ref.~\cite{PezzellaPRD2025}. We show this result in Fig.~\ref{fig:dmqnm} by considering the QNM frequency of a massless scalar field. The cross represents the Schwarzschild case, and the three straight lines have slopes given by $-(\gamma^2-3\gamma-6)/6$, with $\gamma=0$ (solid), $\gamma=(1/2,5/2)$ (dashed), and $\gamma=(1,2)$ (dotted). We find that in the large-$\ell$ limit, the QNM frequencies are redshifted following a straight line on the complex plane whose slope only depends on $\gamma$, as expected from Eqs.~\eqref{OmegaLR_qgamma} and \eqref{lambdaLR_qgamma}. As a consequence, the redshift relation would allow for not only probing the gravity effects contributed by DM halos, but also diagnosing different setups of BH-DM models.

\section{Conclusions and discussions}\label{Conc}

In this work, we have constructed an exact and unified family of static, spherically symmetric BH geometries sourced by generic cored/cuspy DM distributions. The halo is incorporated as an anisotropic matter source with $P_{r}=-\rho$, and the corresponding BH-halo geometry is obtained self-consistently from the Einstein equations. The general ${\alpha,\beta,\gamma}$ density profile encompasses several widely used models, including the NFW, Moore, Hernquist, Jaffe, and core/cusp Dehnen distributions. The resulting geometries are asymptotically flat for \(\beta\geq3\), with a finite ADM mass when \(\beta>3\). Although the total halo mass, or its corresponding normalization for infinite-mass profiles, may greatly exceed the BH mass, the amount of DM enclosed on the BH scale is strongly suppressed by the hierarchy \(M_{\rm BH}\ll r_{s}\). Consequently, the event horizon remains perturbatively close to its Schwarzschild value, and the halo-induced curvature near it can remain parametrically small. The weak energy condition is satisfied for positive, monotonically decreasing density profiles, whereas the strong energy condition is violated near the horizon for cores and shallow cusps with \(0\leq\gamma<2\), is marginally saturated at \(\gamma=2\), and can be satisfied for steeper cusps.

 A central result of our analysis is that the leading strong-field corrections are governed not by the total halo mass alone, but by the dimensionless parameter $q_{\gamma}$, which characterizes the DM contribution on scales $r\sim M_{\rm BH}$.
 In the perturbative regime $q_{\gamma}\ll1$, the leading corrections are determined by the inner logarithmic slope $\gamma$, whereas the dependence on the transition parameter $\alpha$ and the outer slope $\beta$ enters only at higher orders in the near-zone expansion.
 Moreover, in the perturbative regime $q_{\gamma}\ll1$, cuspy profiles with $\gamma>0$ shift the unstable light ring outward, whereas the cored case $\gamma=0$ leaves its radius unchanged; the ISCO, however, moves outward throughout $0\leq\gamma<3$. At the same time, both characteristic orbital frequencies $\Omega_{\rm LR}$ and $\Omega_{\rm ISCO}$ are redshifted, while the critical impact parameter $b_{\rm crit}$ increases and the light-ring Lyapunov exponent $\lambda_{\rm LR}$ decreases. 
The surrounding halo therefore enlarges the geometric-optics capture cross section and lengthens the coordinate-time instability scale $\tau_{\rm LR}=\lambda_{\rm LR}^{-1}$, corresponding to a weaker radial instability of the null circular orbit.

The lensing analysis provides a complementary manifestation of the same dependence on the halo’s inner structure. In the weak-deflection regime, the DM contribution is strongly suppressed for a cored distribution and becomes progressively larger for the cuspy profiles considered here as the inner logarithmic slope increases. A similar ordering emerges in the strong-deflection regime: near the critical photon-capture threshold, cuspy halos enhance the leading logarithmic growth of the bending angle, whereas the cored profile produces no correction to this coefficient at the order considered. The steepest Dehnen cusp therefore yields the largest leading lensing modification among the models examined. This comparison is understood at fixed halo compactness and scale hierarchy, and, in the strong-deflection case, at the same relative distance from each model’s critical impact parameter. Subleading contributions and the finite part of the deflection angle remain sensitive to the complete radial structure of the halo. Weak and strong lensing thus probe complementary regions of the spacetime: weakly deflected rays accumulate the halo’s influence along an extended trajectory, while near-critical photons spend a prolonged time close to the unstable light ring and are primarily sensitive to the DM enclosed in the relativistic region.

The halo imprint also extends to the ringdown spectrum. In the eikonal regime, the oscillation and damping rates acquire distinct environmental shifts whose relative behavior is governed, at leading order, by the inner slope $\gamma$. The scalar-field modes studied here are consistent with this trend and approach the predicted trajectory in the complex-frequency plane as the multipole number increases. 
Thus, within the perturbative regime, the quasinormal-mode correction carries information not only about the overall strength of the halo contribution, but also about its central structure, potentially helping to distinguish cored from cuspy configurations.

Taken together, these results establish a direct connection between the inner structure of galactic DM halos and a coherent set of strong-field observables: the near-horizon geometry, characteristic circular orbits, photon capture, weak and strong lensing, orbital instability, and ringdown. Although these effects remain perturbative within the astrophysical hierarchy considered here, their common dependence on the halo profile and particularly on its inner slope may be more useful than any single correction taken in isolation. The present framework thus offers a useful basis for incorporating galactic environments into precision tests of BH spacetimes, with natural extensions to rotating geometries, more realistic near-horizon matter distributions, a complete analysis of gravitational perturbations and late-time propagation, and direct comparisons with horizon-scale imaging and future high-precision gravitational-wave observations.


\section*{Acknowledgments}
H.H. and S.Z. gratefully acknowledge the University of Bremen for its warm hospitality during the preparation of this work. C.Y.C is supported by the Special Postdoctoral Researcher (SPDR) Program at RIKEN and RIKEN Incentive Research Grant (Shoreikadai) 2025.



\end{document}